\documentclass[%
 reprint,
showpacs,
 amsmath,amssymb,
 aps,
]{revtex4-1}

\usepackage{graphicx}
\usepackage{dcolumn}
\usepackage{bm}

\begin{document}

\title{Effect of a Microscopic Roughness on Biological Adhesion of a Spherical Capsule} 

\author{Aleksey V. Belyaev}
 \email{aleksey{\_}belyaev@yahoo.com}
 \affiliation{ M. V. Lomonosov Moscow State University, Faculty of Physics, 119991 Moscow, Russia}
  \affiliation{ RUDN University, 6 Miklukho-Maklaya str., 117198 Moscow, Russia}

\date{\today}

\begin{abstract}
By means of computer simulations, this work addresses adhesion of a  deformable spherical capsule to a micro-rough surface consisting of a periodic array of pillars. Depending on the micro-relief topography, three different adhesion regimes have been observed: 1) weak adhesion without deformation of the membrane (fakir state); 2) strong adhesion with deformation of the capsule membrane and binding to the bottom wall (nested or contacting state); 3) impalement of the capsule by micropillars. It has been found that a periodic micro-relief implies a favorable positioning of the capsule on rough surfaces. 
\end{abstract}

\maketitle

\section{Introduction}

Surfaces of indwelling medical devices can be colonized by human pathogens that can form biofilms and cause infections. Since these biofilms are usually resistant to antimicrobial therapy it is  necessary to find  efficient ways of the anti-microbal treatment or prevention of biofilm formation \cite{chung2007}. One of the novel strategies for preventing  development of biofilms is to alter the surface properties of biomaterials, including surface topography. Surfaces with natural or artificial micro-/nano-relief have been recognized as a promising way to control the wettability \cite{priezjev.nv:2005, quere.d:2008, quere2005, bocquet2007}, self-cleaning and hydrodynamic properties \cite{barrat:99, belyaev.av:2010b, sphere-pre-2011, Asmolov2015,belyaev.av:2010a,bocquet2007, lauga2003}. Such materials have attracted an attention of researchers, as there is a great expectation that  micro-roughness may help to control the bacterial adhesion. Indeed, a reduced affinity of bacterial cells to superhydrophobic materials leads to a rare surface coverage as demonstrated by recent experiments \cite{Oliveira2012, Ishizaki, Lourenco}. At the same time, controversial data exists suggesting that surfaces with micro-relief may enhance bacterial adhesion and proliferation of cells as compared to ``flat'' surfaces \cite{Shiu, Ishizaki} or even demonstrate a selectivity to different bacterial species \cite{fadeeva2011}. Depending on the surface topography and physico-chemical properties, some materials may demonstrate an anti-microbal effect as a consequence of surface's spiky topography: the impalement of cell membrane by the micro-relief may damage and kill the adherent bacteria \cite{hasan2013, Ivanova2012, Pogodin}.
While the known experimental attempts support this idea in general, there is a certain lack of predictive models in this field \cite{hasan2013}. Therefore, it is important to understand the basic physical and physico-chemical mechanisms responsible for  antibacterial action. In this regard, the phenomenon of biological adhesion should be considered from the molecular point of view and linked to the mechanics of biological cells.

Biological adhesion is ubiquitous in nature. Adhesive forces comprise numerous physical processes including Van der Waals forces between cell membranes, lubrication hydrodynamic forces, electrostatic interactions and electrokinetics, etc. However, it has been recognized that an important and reliable contribution comes from the key-lock interactions between cell adhesion molecules (CAMs), also known as the receptors, and their molecular ligands or a substrate. The substrate could be another ligand protein, or surface chemical groups of the implant, or both. While the non-specific interactions are more or less well understood \cite{Seifert2007}, the kinetics and mechanics of key-lock binding is more challenging.

 Most of the contemporary approaches rely on the idea that bond formation and breakage could be viewed as a statistical process modelled as a reversible chemical
reaction between populations of free membrane receptors and their ligands \cite{ModyKing2008-2, Dembo1988, HammerApte1992}. Most of the CAMs belong to four protein
families: the immunoglobulin superfamily, the integrins, the cadherins, and the selectins.
Some CAMs demonstrate high selectivity and require a specific
ligand to attach, while others are less selective and may interact with many ligands and materials.

Many features of cell-to-cell and cell-to-substrate adhesion
have been studied using living cells. But
due to the complexity of the cell, it is often difficult to achieve the exact same experimental conditions reproducibly \cite{Seifert2007}. Therefore, giant liposomes, or vesicles, have been proposed as a model system for experimental and theoretical studies of biological adhesion. Such artificial vesicles reproduce well the mechanical properties of the cell membrane \cite{Vezy, Irajizad, Seifert2007, Abkarian}. CAMs may be introduced into vesicles to model ligand-receptor interactions \cite{Sackmann, Seifert2007, Fenz}. 

 Vesicle adhesion has been a subject of an active research \cite{Evans1980, SeifertLipowsky1990, Fenz, Deserno2007, Smith2004}. A number of studies addressed the ligand-receptor adhesion of cells and vesicles to flat surfaces \cite{korn2008, Dembo1988, HammerApte1992, luo2016, Seifert2007, Irajizad, Firrell, Abkarian, Khismatullin}. However, when it comes to anti-bacterial micro-patterned surfaces, the interplays between a micro-relief and cell elasticity should be taken into account. Another issue of a practical interest is the design of hemo-compatible and anti-thrombotic surfaces for  implants and microfluidics. 
 
 The present work theoretically studies the adhesion of a deformable spherical capsule to a surface with a periodic micro-relief in a steady fluid. The structure of the paper is as follows. First, the description of the computer model is presented. Then, equilibrium shapes of a capsule on a flat substrate are studied for different values of model parameters and compared to existing experimental data. After that the simulation results for the adhesion on a micro-rough surface consisting of a periodic array of pillars is presented, and a state diagram regarding the geometry of the surface relief has been plotted. The discussion of the results concludes the paper.

\section{Materials and Methods}

 \subsection{Problem setup}
\begin{figure}
 \includegraphics[width=\columnwidth]{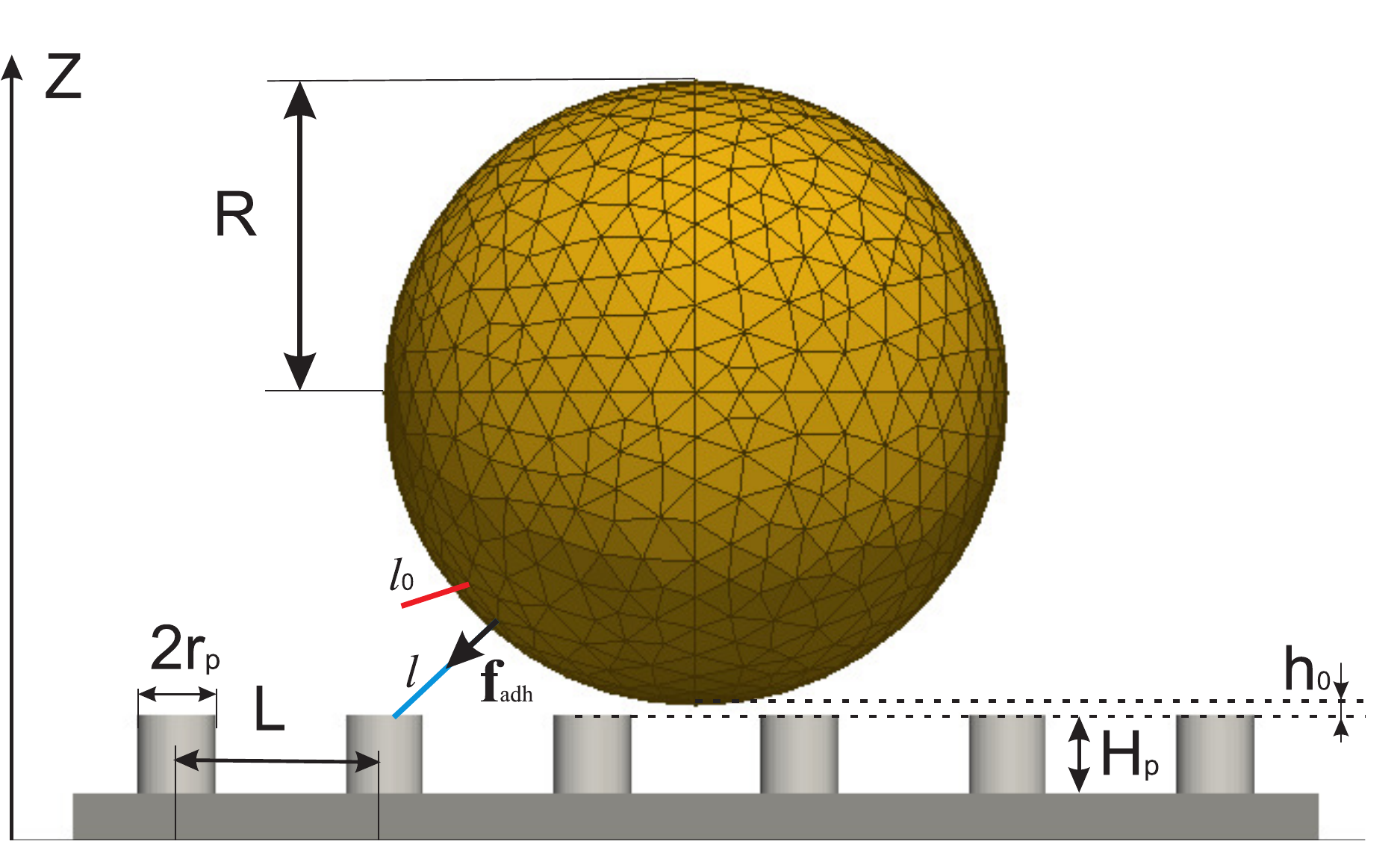}
  \caption{  The initial positioning of the capsule near a periodic array of pillars. The initial gap width between the capsule and tops of the pillars is equal to $h_0$. The capsule radius $R$, the period of pillars is $L$, pillar radius $r_p$. The adhesive ligand-receptor bonds may form between the capsule and the pillared surface. The resting bond (red) of length $l_0$ exerts no force on the capsule (not shown further). The active bond of a length $l>l_0$ (blue) exerts the force $\textbf{f}_{\rm adh}$ of the capsule's Lagrangian surface point.}
\label{overview}
\end{figure}

Consider a micro-rough surface consisting of a periodic micro-pillar array, Fig \ref{overview}. A spherical capsule is allowed to form adhesive bonds (i.e. elastic bridges ) between its membrane and the micro-rough surface.
 The system's behaviour is governed by the formation of adhesion bonds between the capsule and surfaces, capsule elasticity and viscous forces from the surrounding fluid.  
 
A three-dimensional computer model was used to study bioadhesion of a soft spherical capsule of radius $R=6$ $\mu$m to several micro-rough (pillared) surfaces. The setup consisted of a bottom surface with a periodic pillar-shaped roughness. The rough bottom wall was represented by a geometrical constraint consisting of rhomboid and a periodic array of cylinders.  The pillars had circular cross-section, their radius and height were altered during the study, and  individual simulations were performed for each set of geometric parameters. Numerical simulations have been carried out in a rectangular box of $b_x \times b_y \times b_z$ sizes with $b_z=32$ $\mu$m; $b_x$ and $b_y=b_x$ were chosen for each simulation so that the box contained 6 periods of pillars in $x$ and $y$ directions, Fig. \ref{overview}.

The capsule was modelled as a deformable sphere. It was placed near the bottom surface at a distance $h_0$ before each simulation run. Normally the distance between the sphere and the surface was set to $h_0=0.2$ $\mu$m to ensure the adhesive bond formation. The capsule was able to form adhesive ligand-receptor bonds with every point of the rough surface. The whole system has been immersed into a viscous Newtonian fluid with certain properties (density $\rho = 1.0$ g/cm$^3$, kinematic viscosity $\nu=1.5$ mm$^2$/s). The no-slip hydrodynamic condition has been imposed on the rough bottom surface. The top plane limiting the simulation box at $z=32$ $\mu$m was impenetrable for the capsule, and the no-slip hydrodynamic boundary condition has been imposed there.

 \subsection{Numerical method}
The computer simulations were based on a combination of the Lattice Boltzmann method (LBM) \cite{Succi} with the Lagrangian Particle Dynamics (LPD) implemented in ESPResSo open-source software (ver. 3.4) \cite{DunwegLadd,Cimrak2012278}.  The software package has been modified in order to implement the Monte-Carlo model of adhesive bonds kinetics \cite{ModyKing2008-2, HammerApte1992}. 

LBM is used as a fast solver for hydrodynamic equations, that inherits from lattice gas automata simulations. The method rests upon the Boltzmann's kinetic equation that describes spacial-temporal changes of a one-particle distribution function $f(\textbf{x}, \textbf{u}, t)$  \cite{Succi}.  A discretization scheme D3Q19 has been used in this work, i.e. the fluid is treated as packets of fluid particles moving from one node to a neighbouring node of a 3-dimensional periodic cubic grid in 19 possible directions. A regular mesh of Eulerian spacial sites $\{\textbf{x}\}$ and lattice velocities $\{\textbf{c}_i\}$, this function $f_i(\textbf{x}, t)$ obeys the following equation:
\begin{equation}
    f_i(\textbf{x}+\textbf{c}_i \Delta t, t+ \Delta t) = f_i(\textbf{x}, t) + \Omega_i(\textbf{x}, t),
\end{equation}
and the evolution of the system could be found by consequent iterations. Here $f_i(\textbf{x}, t) \equiv f(\textbf{x}, \textbf{c}_i, t)$ and $\Omega_i(\textbf{x}, t)$ is the collision operator that defines a rheology of the system. For a viscous incompressible fluid a single-relaxation time approximation (or Bhatnagar-Gross-Krook approximation) has been proven to precisely reproduce the hydrodynamics for low Reynolds and Mach numbers \cite{Bhatnagar1954, Reasor2012, Mountrakis2015}. As in the present work the fluid medium is needed to account for viscous damping and drag on the adhering microscopic capsule, the BGK approximation was an appropriate choice:
\begin{equation}\label{OmegaLB}
    \Omega_i(\textbf{x}, t) = -\frac{1}{\tau}(f_i(\textbf{x}, t) - f_i^{\rm eq}(\textbf{x}, t)).
\end{equation}
The equilibrium distribution function $f_i^{\rm eq}(\textbf{x}, t)$ corresponds to the series expansion of Maxwell-Boltzmann distribution for small velocities \cite{Bhatnagar1954}. 
 A single relaxation time $\tau = 0.5 + \nu /(c_s^2 \Delta t)$ has been used for the collision step, where $c_s^2 = (1/3)(\Delta x/\Delta t)^2$ is the lattice speed of sound.
For setting up the no-slip hydrodynamic boundaries the ``link bounce back'' method has been used\cite{DunwegLadd}.

A deformable spherical vesicle (a capsule) was represented by a triangular mesh of 727 Lagrangian surface points (or LSPs) connected in triangles by unbreakable elastic neo-Hookean springs. Particle dynamics approach has been used to simulate the motion of LSPs, according to which the position $\textbf{r}_{\rm LSP}$ and the velocity $\textbf{v}_{\rm LSP}$ of each LSP are found from the solution of the following differential equations: 
\begin{equation}
   \frac{d \textbf{v}_{\rm LSP} }{d t} = \frac{ \textbf{F}}{m}, \quad \frac{d \textbf{r}_{\rm LSP} }{d t} = \textbf{v}_{\rm LSP},
\end{equation}
where $ \textbf{F} = \textbf{F}_{\rm elast} + \textbf{F}_{\rm visc} + \textbf{F}_{\rm adh} + \textbf{F}_{\rm rep}$ is the total force exerted on the LSP. The viscous Stokes-like force $\textbf{F}_{\rm visc}$ is used for the coupling between the particles and the fluid, and $\textbf{F}_{\rm elast}$ accounted for the elasticity of the capsule. The adhesive force is given by a vector sum over all  adhesive bonds active at this moment of time $\textbf{F}_{\rm adh} = \sum_{\rm bonds} \textbf{f}_{\rm adh}$. The last term $\textbf{F}_{\rm rep}$ corresponds to a short-range soft-sphere repulsion force to avoid a non-physical penetration of LSPs into the adhesive substrate.

The coupling between the fluid and the particles has been provided via the viscous-like force applied to the membrane surface points with respect to the local fluid velocity \cite{DunwegLadd, Cimrak2012278}. This force, by the analogy with the Stokes formula, was proportional to the difference of the velocity $\mathbf{v}_{\rm LSP}$ of the LSPs and the local fluid velocity $\mathbf{u}$ derived from the neighboring LB lattice nodes, $\mathbf{F}_{\rm visc} = \xi (\mathbf{u}-\mathbf{v}_{\rm LSP})$. The opposite force $-\mathbf{F}_{\rm visc}$ was transferred back to the fluid.

The model of the spherical capsule accounted for stretching elasticity, bending rigidity, conservation of volume and surface \cite{Cimrak2012278}: $\textbf{F}_{\rm elast} = \mathbf{F}_{\rm sp}+ \mathbf{F}_{\rm b} +  \mathbf{F}_{\rm a} + \mathbf{F}_{\rm v}$.
 The elastic force applied to
the membrane points in the case of stretching (compression) of mesh edges is given by neo-Hookean law:
\begin{equation}
   \mathbf{F}_{\rm sp} = k_s \frac{\lambda^{0.5} + \lambda^{-2.5}}{\lambda+\lambda^{-3}}\frac{\Delta l}{l_0} \mathbf{n} ,
\end{equation}
where  $\Delta l = l-l_0 $ is the spring elongation relative to its equilibrium length $l_0$, $\lambda= l/l_0$, $k_s$ is the stretching spring constant, and $\mathbf{n}$  is the unit vector pointing from one membrane point at another.

The bending elasticity force was meant to provide equilibrium angle $\theta_0$ between two adjacent mesh triangles:
\begin{equation}
   \mathbf{F}_{\rm b} = k_b \frac{\Delta \theta}{\theta_0} \mathbf{n}_b ,
\end{equation}
where $\mathbf{n}_b$ is the unit vector normal to the triangle (pointing at the exterior of the capsule),  $\Delta \theta$ is the angle deviation from $\theta_0$,  $k_b$ is the bending elasticity constant. This force was applied to the
vertex not belonging to the common edge of adjacent triangles. The opposite force divided by two was applied to the two vertices lying on the common edge.

The surface conservation force applied to the mesh nodes for maintaining the surface area of each mesh triangle was
\begin{equation}
   \mathbf{F}_{\rm a} =  - k_{\rm al} \frac{\Delta S_i}{(S_i^0)^{0.5}} \mathbf{w} - k_{\rm ag} \frac{\Delta S_g}{S_g^0} \mathbf{w},
\end{equation}
where $\Delta S_i= S_i-S_i^0$ is the change of the i-th mesh triangle area, $\mathbf{w}$ is a unit vector pointing from the centroid of the triangle at the vertex, $k_{al}$ and $k_{ag}$ are the coefficients.
Finally, the volume conservation force is given by
\begin{equation}
   \mathbf{F}_{\rm v} = - k_v \frac{\Delta V}{V_0} S_i \mathbf{n}_b ,
\end{equation}
where $\Delta V = V - V_0$ is the capsule volume change, $k_v$ is the coefficient of volume conservation. The force $\mathbf{F}_{\rm v}$ has been calculated for each i-th triangle with area $S_i$ and has been evenly distributed over the vertices of this triangle.

Validation tests were based on theory of Goldman et al. for a near-wall sphere motion in a shear flow \cite{Goldman1967} and  Jeffery's orbiting of a freely suspended spheroid in an unbounded shear flow \cite{Jeffery}. The validation results were published earlier \cite{BelyaevPlos2017, BelyaevPRE2018} and they have shown a good conformity with the theory.

\subsection{Kinetics of adhesive bonds}
The ligand-receptor adhesion is mainly determined by the mechano-chemical response of adhesion bonds between cells and surfaces\cite{HammerApte1992}.
In the model, the ligand-receptor binding  have been  modelled
as a reversible chemical reaction between the populations
of free membrane receptors (or cell adhesion molecules, CAMs) and their ligands. These reactions could
be characterized by forward $k_{\rm on}$ and reverse $k_{\rm off}$ reaction rates.

Bell was first to invent the kinetic theory of key-lock binding of living cells and to propose a constitutive relation
between the dissociation rate $k_{\rm off}$ and a force on the bond \cite{Bell}. Following these pioneering works, we use the following expression in the present work:
\begin{equation}
     k_{\rm off} = k_{\rm off}^0 \exp(- f_{\rm adh}/f_D)
\end{equation}
where $k_{\rm off}^0$ is a constant unloaded bond rupture rate.
Here $f_{\rm adh}= | \textbf{f}_{\rm adh}|$ and $\textbf{f}_{\rm adh} = \kappa (l-l_0) \textbf{n}$ is the bond tension force, $\kappa$ is the bond stiffness, $l$ is the stretched bond length, $l_0$ is the non-stretched length of a receptor/microvillus, and $f_D$ is the typical bond tension force required for dissociation. Typical values of $l_0$ for white blood cells is 0.2-0.5 microns \cite{Li2016, Dembo1988, Khismatullin, McConnell}. For blood platelets the equilibrium length may be even greater due to long ligand proteins \cite{BelyaevPRE2018, Jedi, Schneider2007}. In the presented model $f_D=0.01$ nN, as it seems to be comparable in order of magnitude with the bond-breaking force measured in experiments \cite{Thomas2008, Kim2010}.

Dembo et al. \cite{Dembo1988, Dembo1994} have proposed a modification to Bell's original formula, including the exponential ratio of elastic energy to $kT$, instead of forces, however the general pattern of the dissociation rate increasing with bond tension remains.
Constitutive equations of the similar exponential force-dependent form have also been proposed for the on-rate \cite{Bell1984, Piper1998}. It is assumed that binders attach to the surface at a rate that depends solely on the distance the receptor heads have to traverse in order to stick to the surface-immobilized ligand\cite{Mani2012}. Thus the binding is treated as a thermally-activated process dependent on the diffusion constant of the binder head. Following prior works\cite{Mani2012}, here we use the simplest expression for bond formation rate:
\begin{equation}
  \label{Kon}
     k_{\rm on} = k_{\rm on}^0 \exp[- \kappa (r-l_0)/f_D], \quad r<r_{\rm max},
\end{equation}
and $k_{\rm on}=0$ for $r \ge r_{\rm max}$ 
Here $r$ is the shortest distance between a mesh node of a capsule (i.e. a receptor) and an adhesive site on the rough bottom wall.   According to the Bell's theory \cite{Bell}, the on-rate $k_{\rm on}^0 = k_{+} n_l n_r$ is proportional to an intrinsic rate $k_{+}$ of the formation of a ligand-receptor complex and a surface density $n_l$ of ligands/adhesive sites on the bottom surface available for binding; $n_r= 100$ is the number of receptors per LSP. Typical values found in literature \cite{Evans2007,Khismatullin,Bell,Bell1981, Li2016} : $k_{+} =1-100$ $\mu$m$^2$s$^{-1}$, $n_l=100-1000$ $\mu$m$^{-2}$. 
The surface density of ligand molecules on the surface was assumed to be large, compared to the number of active bonds per unit area. Instead of introducing the adhesive sited explicitly, the mean value of adhesion on-rate $k_{\rm on}^0$ was used and it was assumed to be uniform over the bottom rough wall. Here, a new bond can be established between the capsule's LSP and the nearest wall point. The coordinates of this wall point are then memorized until the bond rupture. This approach makes the adhesion kinetics independent from the grid resolution and the initial seeding of the adhesion sites on the rough wall. The surface density of adhesive sites $n_l$ (and thus the affinity of receptors to the wall) can be varied by changing $k_{\rm on}^0$. 

A probabilistic Monte Carlo-like approach \cite{ModyKing2008-2} has been employed to model the ligand-receptor adhesion, and the probabilities of bond formation and dissociation during iteration step $\Delta t$ are given as follows:
\begin{equation}
    P_{\rm on} = 1 - \exp(-k_{\rm on} \Delta t)
    \label{Pon}
\end{equation}
\begin{equation}
    \label{Poff}
    P_{\rm off} = 1 - \exp(-k_{\rm off} \Delta t).
\end{equation}
An additional constraint on the bond length was used that the maximal length of a stretched receptor-ligand complex should not exceed $r_{\rm max}=1.6$ $\mu$m, corresponding to maximal length of a cell's microvillus found in literature \cite{McConnell}. If the bond length exceeded this value, the bond would be assumed inevitably broken.

\subsection{Parameters}

For convenience, the dimensionless values have been used in the following text. The scales for length, force and time were chosen so that $[l]=1$ $\mu$m, $[F]=1$ nN, $[t]=(k_{+} n_l n_r)^{-1}= 1$ $\mu$s. The iteration step for molecular dynamics $\Delta t = 10^{-1}$ $[t]$ was equal to the LB timestep.
The dimensionless elastic coefficients have been chosen  the following way: $k_{\rm al}=k_{\rm ag}=10.0$, $k_{v}=10.0$, $k_b=0.01 k_s R$, where $R=6.0$  is the capsule radius. The coefficient $k_s$ was altered in order to reveal the influence of capsule stiffness on adhesive properties.
The parameter of friction was set to $\xi=0.25$ for a mesh consisting of 727 LSPs in order to obtain the correct drag force on the spherical particle in an unbounded constant velocity Stokes flow, according to the calibration procedure described  earlier \cite{BelyaevPRE2018, Cimrak2017}. The mass of each LSP point was set to $m=10.0$, following Ref.\cite{Cimrak2012278}.
 
The resolution of Eulerian grid nodes for Lattice Boltzmann module 
was set to $\Delta x=1.0$ for most of the simulations. The tests had been performed to ensure the independence of the results from the grid resolution by reducing the spacing to $\Delta x=0.5$ before the simulations. No significant difference was observed regarding the shape of the adherent capsule. However, for thinner pillars $r_p < 0.5$ the lattice unit $\Delta x$ had to be reduced to 0.25 in order to resolve the hydrodynamic boundary.  

\section{Results}

  The ligand-receptor adhesion of a spherical capsule has been studied in a steady fluid, i.e. no external agitation of the fluid far from the capsule was imposed. The final (steady) position of the capsule and its equilibrium shape are referred as ``adhesion state'' in the following text.

   \subsection{Contact angle on a flat surface.~~}

\begin{figure*}
 \includegraphics[width=1.5\columnwidth]{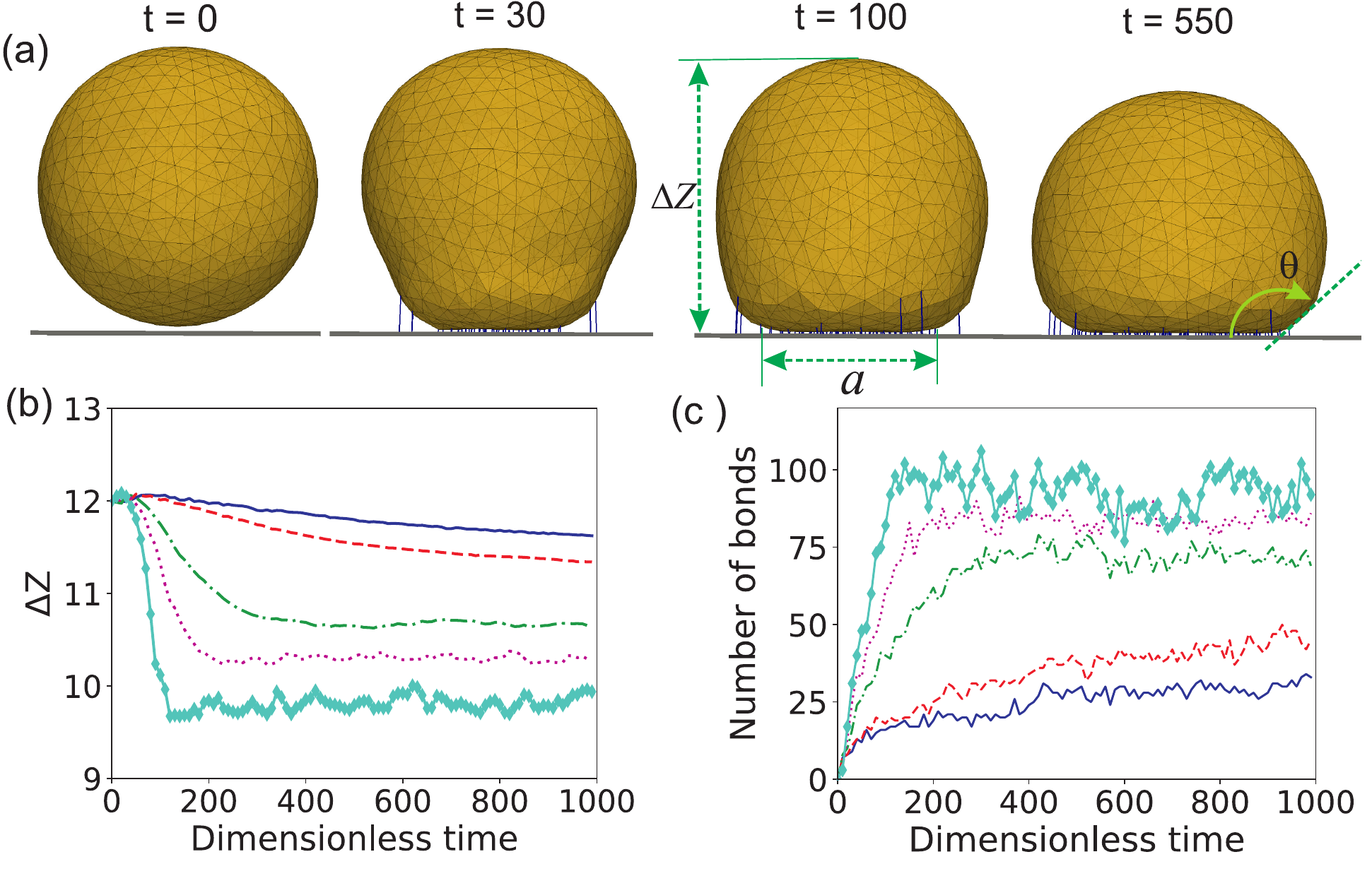}
  \caption{ Ligand-receptor adhesion to the flat surface. (a) Typical snapshots from the initial moments of the simulation. The time is indicated (in units of [t]) above the frames. The contact angle is marked with the green arrow. (b) The height of a capsule $\Delta Z$ versus dimensionless time obtained in the simulation: $\kappa = 0.5$ (the solid blue line), $1.0$ (the dashed red), $5.0$ (the dash-dot green), $10.0$ (the dotted magenta), $20.0$ (the cyan line with symbols); $k_s=1.0 $, $k^0_{\rm on} = 1.0$, $k^0_{\rm off}=0.01$, $l_0=0.3$. (c) The number of adhesive bonds versus time for the same parameters as in the panel (b). }
\label{Fig-flat-1}
\end{figure*}

\begin{figure}
 \includegraphics[width=\columnwidth]{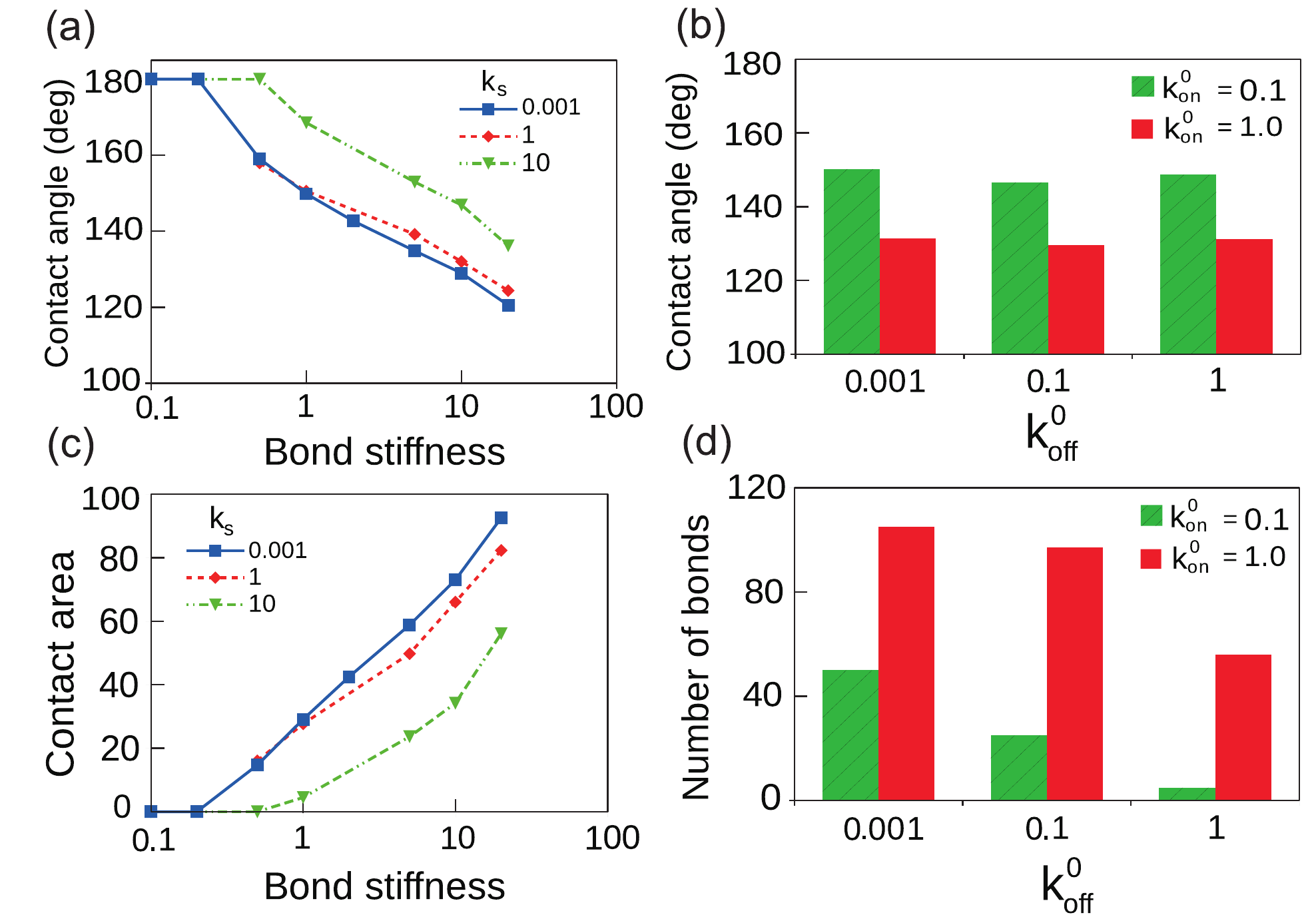}
  \caption{ Adhesion to the flat surface. (a) The dependence of the contact angle between a capsule (vesicle) and a flat adhesive surface for different values of the bond stiffness $\kappa$ and the membrane shear modulus $k_s$. (b) The contact angle between the capsule and the wall as a function of bond association and dissociation constants $k_{\rm on}$ and $k_{\rm off}$. (c) The dependence of the adhesive contact area $A_{\rm adh}=\pi a^2/4$ for different values of the bond stiffness $\kappa$ and the membrane shear modulus $k_s$. Here the contact diameter $a$ is defined as shown in Fig.\ref{Fig-flat-1}(a). (d) The number of adhesive bonds at the end of simulations as a function of bond association and dissociation constants. Here $l_0=0.3$; in panels (a) and (c) $k^0_{\rm on} = 0.1$, $k^0_{\rm off} = 0.01$; in panels (b) and (d) $k_s=0.01$ and $\kappa = 1.0$.}
\label{Fig-flat-2}
\end{figure}

The first set of simulation has been carried out with a flat surface.
In all simulations capsules reached the equilibrium adhesive state within a timespan $1000 \times [t]$ from the start. After the initial adhesive bond formation the capsule has been pulled to the surface and deformed by the simultaneous action of adhesive and wall repulsion forces, Fig.\ref{Fig-flat-1}. As the capsule was pulled to surface the new bonds were formed. The pulling to surface behavior is the consequence of the new bonds emerging in the pre-stretched state according to Eq.(\ref{Kon}). The capsule resembled a liquid drop, hence it was convenient to characterize the adhesiveness in terms of a contact angle. The contact angle on a flat surface has been measured in the first set of simulations by analysing the geometry of the capsule in Paraview software. Depending on the stiffness of the bonds, capsule shear modulus $k_s$ and non-stretched bond length $l_0$, the value of the contact angle changed within a range from 180 degrees for a stiff capsule (when capsule was not deformed at all) to 120 degrees for a deformable one, Fig.\ref{Fig-flat-2}. 

The shapes of the capsule on a flat surface, that were obtained in the simulation, are qualitatively similar to those reported in experimental works for vesicles \cite{Abkarian, Vezy, Irajizad} and white blood cells \cite{Firrell}. The contact area between a resting capsule and the flat bottom wall was measured in the simulations as $A_{\rm adh}=\pi a^2/4 = 40-100$ $\mu$m$^2$, depending on the bond stiffness $\kappa$ and capsule elasticity $k_s$. This range coincides almost exactly with experimentally observed values for vesicles of the radius $R \approx 6$ microns \cite{Irajizad}. The value $A_{\rm adh}=14$ $\mu$m$^2$ for $k_s=10.0$ and $\kappa =1.0$ (Fig.\ref{Fig-flat-2}(c)) is close to the value of $12$ $\mu$m$^2$ for a resting leukocyte in a steady fluid \cite{Firrell}. These facts support the validity of the computer model.

   \subsection{Effect of surface roughness.~~}

\begin{figure}
 \includegraphics[width=\columnwidth]{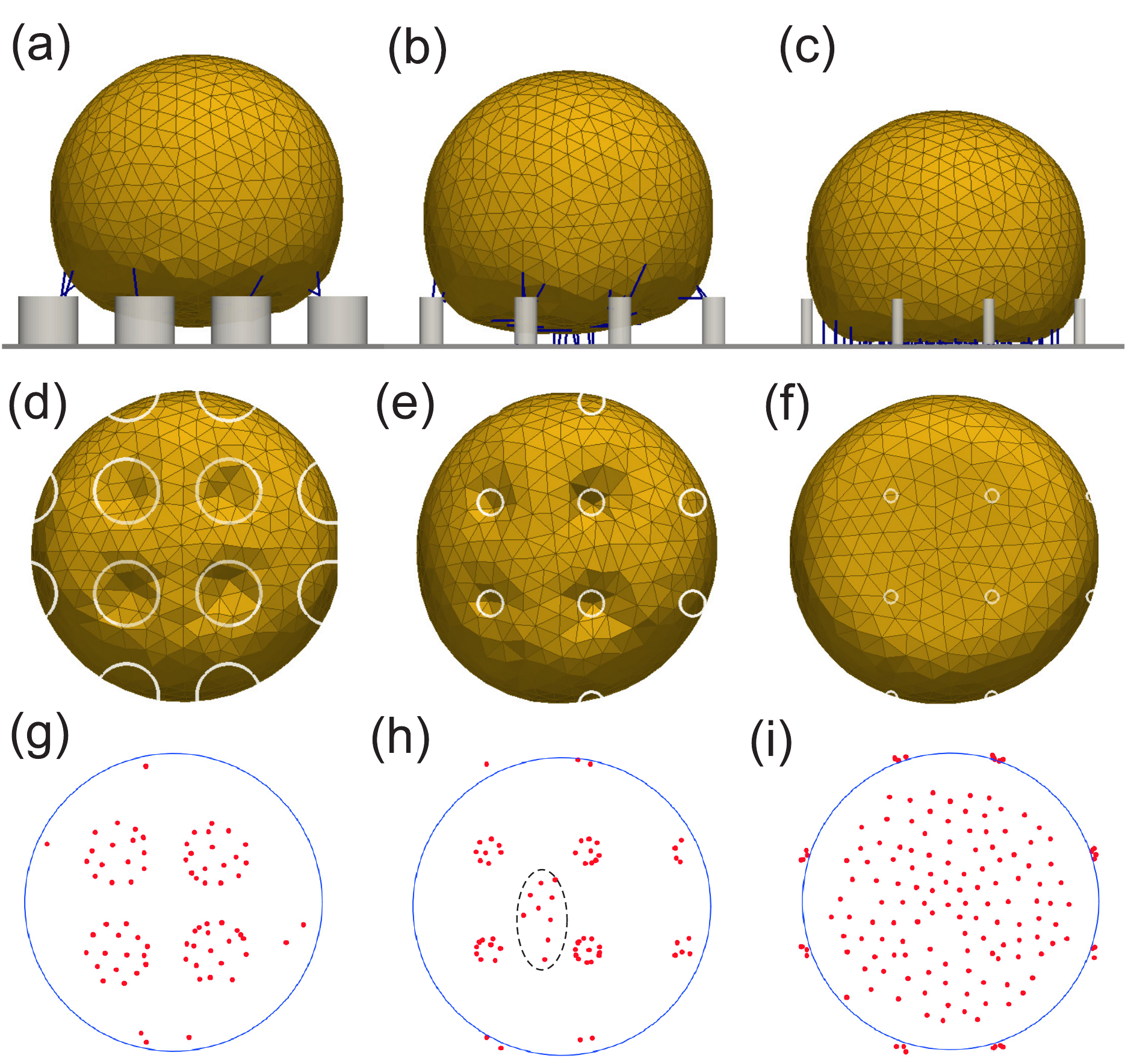}
  \caption{ Adhesive states of the capsule on a pillar-shaped roughness. By changing the pillar diameter $r_p$ one can observe three different adhesive states: fakir (a,d,g), nested (b,e,h) and impaled (c,f,i). (a-c) The side view of the simulation, where the capsule has reached its steady adhesive state. (d-f) The bottom view of the simulation with white circles marking the positions of the pillars. The deformation of the capsule is visible. (g-i) The map of the binding points on the surface, view from the bottom of the capsule in the same projection as (d-f).}
\label{Fig-rough-3}
\end{figure}

In particular, three adhesion states have been  observed on a rough  surface, Fig.\ref{Fig-rough-3}. Firstly, the \textit{`fakir'} state has been found similar to the Cassie-Baxter state of water drops on a superhydrophobic surface \cite{quere2005}. In this case the capsule formed adhesive bonds only with pillars, but not with the bottom (base) surface of the rough wall. For this state to manifest itself the distance between the pillars has to be smaller than a pillar radius. Also,  elasticity of the capsule should prevail over the adhesive bond tension pulling the capsule's membrane towards the rough wall.
  Secondly, the \textit{nested} (or \textit{contacting}) state has been observed when the capsule was able to form adhesive bonds with the bottom of the texture not pierced by the pillars. In this case the elasticity of the capsule allowed it to squeeze between the pillars without breaking through the  membrane. This state has been observed with the pillars not very high and their radius not too small to keep the balance between adhesive forces and the  elasticity.
  Finally, the \textit{impaled} state with the  membrane  pierced by the pillars has also proved possible with the pillars fine enough to pass between the surface mesh nodes (LSPs) and penetrate into the capsule's interior. The  impaled and nested states are somewhat similar to the Wenzel state of water drops on rough surfaces \cite{quere2005, quere.d:2008}. However, here in a case of a capsule (or a vesicle/cell) we can distinguish between nested and impaled states depending on whether or not the capsule has retained its integrity. The latter situation could be associated with critically damaged membrane inconsistent with the cell's normal functioning and living. In principle, this property  could be used for the design of antimicrobial surfaces.

   \subsection{Adhesion as a wetting phenomenon.~~}

\begin{figure}
 \includegraphics[width=\columnwidth]{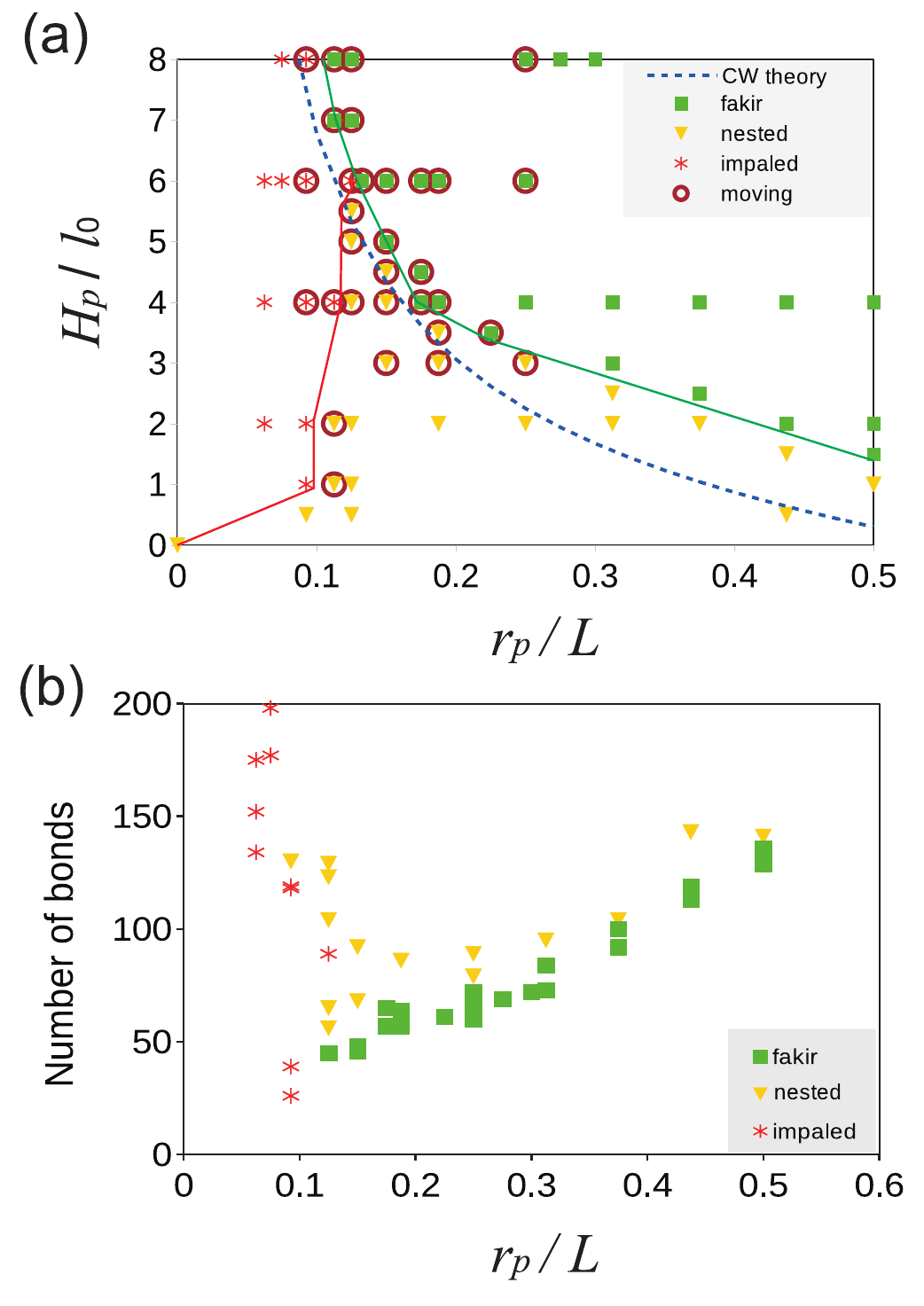}
  \caption{ (a) Adhesion diagram for a deformable capsule near a micro-rough surface. Only geometry of the bottom wall was altered in this set of simulations, while the properties of the capsule and adhesive bonds remained the same. The vertical axis corresponds to $H_p/l_0$ and the horizontal axis - to $r_p/L$. The dashed line corresponds to the Cassie-Wenzel transition theory for a liquid drop on a rough surface, Eq.(\ref{CWtheor}). The green circles correspond to the fakir state, yellow triangles - to the nested (contacting) state and the red asterisks - to the impaled state. The brown circles denote the simulations, during which the cell moved from its initial position in the lateral direction to reach the equilibrium position pillar-centred. (b)The number of  adhesive bonds between the cell and the surface in the steady adhesive state as a function of the surface geometry parameter $r_p/L$. Here the following parameters have been used: $\kappa/k_s = 100$, $k_s=0.01$, $l_0=0.5$, $k_{\rm on} = 1.0$ and $k_{\rm off} = 0.01$, $R=6$}
\label{Fig-state-diag}
\end{figure}

The adhesion state depends on the micro-relief geometry. Fig. \ref{Fig-state-diag} summarizes the results of simulations. We can see that in general the impaled state manifests itself in a case of very thin pillars (small $r_p<0.1$). To the contrary,  wide and long pillars lead to the fakir state with less adhesive bonds and insignificant deformations of the cell. The intermediate nested state is observed when the pillar height $H_p$ does not exceed 4-5 $l_0$.

Here an analogy between wetting and  ligand-receptor adhesion may be stated \cite{Sackmann, Seifert2007}. Let us compare the simulation results with the wetting transition criterion. The transition between a completely wetted (Wenzel) and a non-wetted (Cassie-Baxter) states for water on a rough surface is given by the following formula \cite{CWtransition}:
\begin{equation}
    \frac{\phi-1}{f-\phi} = \cos\theta_0
\end{equation}
where $\phi=\pi r^2_p/L^2$, $f =1+2\pi r_p H_p/L^2$ is the surface roughness,  $\theta_0$ is a contact angle on the flat surface. From this expression for the pillared periodic surface it follows that:
\begin{equation} 
    \frac{H_p}{l_0}\frac{l_0}{L}=\frac{(\phi-1)(1+\sec\theta_0)}{2(\pi\phi)^{1/2}}
    \label{CWtheor}
\end{equation}
In Fig.\ref{Fig-state-diag}(a) the dashed line corresponds to Eq.(\ref{CWtheor}). The line differentiates the fakir state (which is analogous to Cassie-Baxter wetting state) from the nested and the impaled states (both similar to Wenzel state). However, a discrepancy could be noted and attributed to i) the strict membrane area conservation condition in our model; ii) non-equilibrium transitions observed in simulations in contrast to the assumed equilibrium transition in theory.

     \subsection{Crawling to the equilibrium position.~~}

\begin{figure}
 \includegraphics[width=\columnwidth]{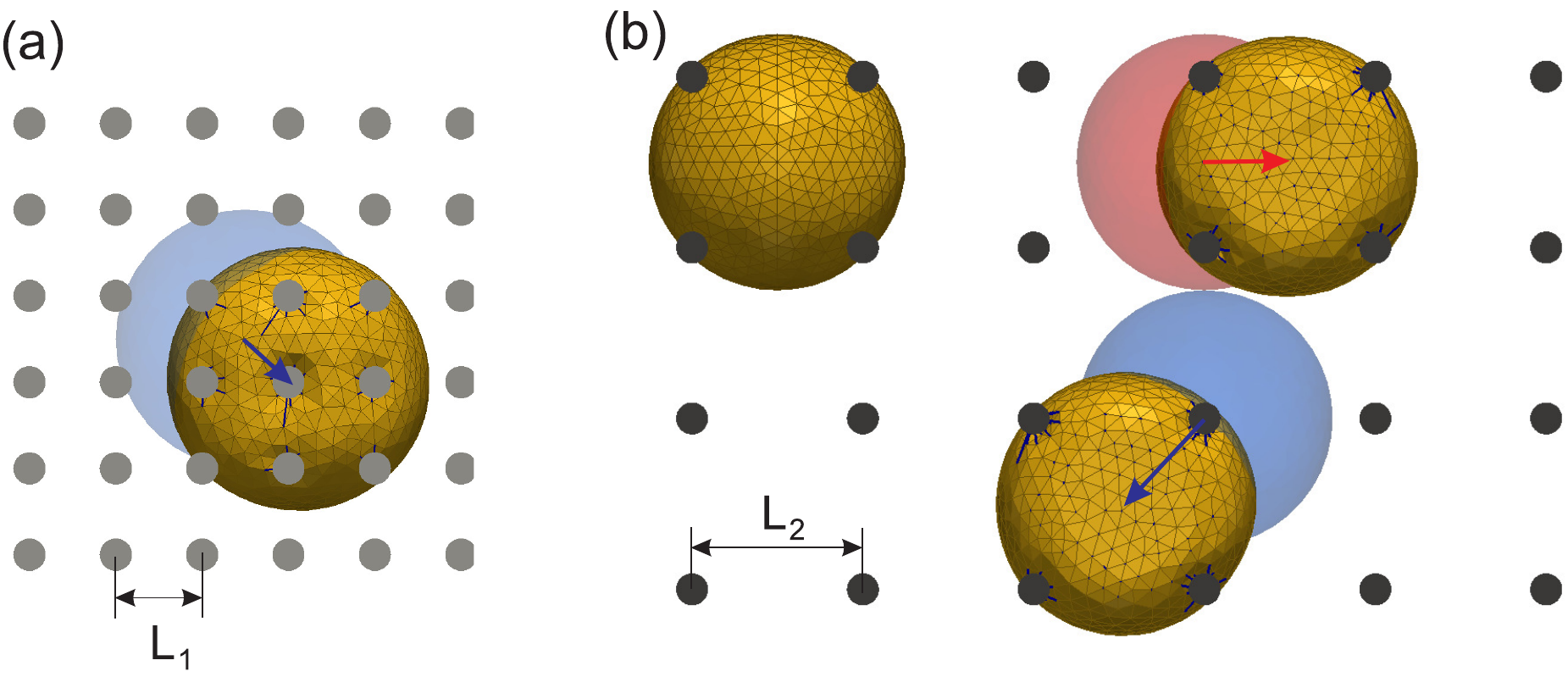}
  \caption{ The dependence of the equilibrium (final) position of the capsule on the period the the micro-roughness in the steady fluid. Here the periods of the pillars are $L_1=4$ (a) and $L_2=8$ (b). For a relatively dense micro-relief (a) the pillar-centered position was more favourable than the initial gap-centered placement. For a sparser micro-roughness (b), the capsule tends to rest between the pillars, thus demonstrating the gap-centered position in the end of simulation for any initial placement. The red and blue circles represent the initial positions of the capsule, and the arrows demonstrate its displacement during time. Each simulation run was independent, and in panel (b) the results of three simulation runs are combined into one figure solely for the representation. The parameters used in this example are as follows:  $\kappa/k_s = 100$, $k_s=0.01$, $l_0=0.5$, $k_{\rm on} = 1.0$ and $k_{\rm off} = 0.01$. }
\label{Fig-crawl}
\end{figure}

\begin{figure}
 \includegraphics[width=\columnwidth]{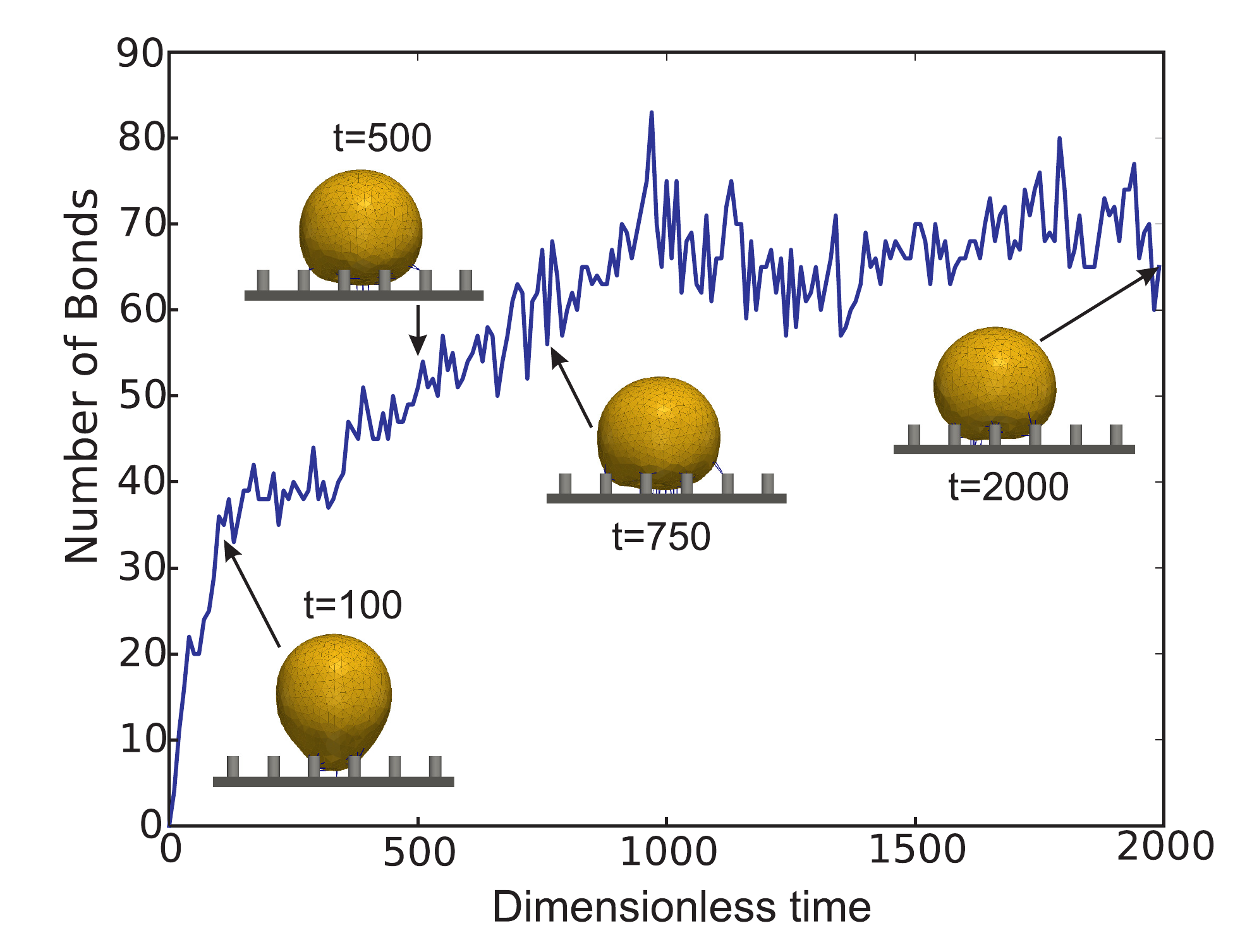}
  \caption{ The evolution of the number of bonds during the simulation time. The small panels in the main plot show  capsule shapes at  corresponding moments of time.  Here the period of the pillars $L=4$, height of the pillars $H_p = 2$, pillar radius $r_p/L=0.15$, the non-stretched bond length $l_0=0.5$,  $\kappa/k_s = 100$, $k_s=0.01$, $k_{\rm on} = 1.0$ and $k_{\rm off} = 0.01$. }
\label{Fig-crawl-bonds}
\end{figure}
 
In the majority of  simulation runs (denoted by brown circles in Fig.\ref{Fig-state-diag}) the capsule crawled form its initial  position in plane $xOy$ to a more favourable position. For the pillar period $L=4.0$ and the capsule's radius $R=6.0$ this equilibrium (final) location was above one of the nearest pillars. Let us call it as a ``pillar-centered'' position in contrast to the ``gap-centered'' initial placement of the capsule. The crawling effect has been observed in the simulation for both fakir and nested adhesion states (see Supplementary Video 1). The crawling could also be the cause of a spontaneous transition from the fakir state to the impaled state if the pillars are narrow, as observed in the simulation for $H_p/l_0=6.0$ and $r_p/L=0.125$ on Fig.\ref{Fig-state-diag}(a). For this case, the initial fakir state was metastable, so that the crawling in lateral direction destabilized it and led to the impalement of the membrane by the pillars.

The main factor influencing the final (equilibrium) cell's location is the ratio of the period $L$ of the pillars to the cell's size $R$. For the sparse pillar array $L>R$ the gap-centered location is more favourable, while for the dense pillar array $L<R$ the pillar-centered position was stable, Fig.\ref{Fig-crawl} (see also Supplementary Video 2). On the flat surface no crawling has been observed in the model. Another observation is that the reduction of the maximum bond stretch distance to the value $r_{\rm max}= 0.8$ suppressed the crawling in the simulations. This suggests that the pre-stretched non-equilibrium far-reaching bonds are responsible for the observed destabilisation of the initial position of the capsule.  

The proposed mechanism of this crawling is as follows. As the initial gap-centered lateral location of the capsule (for $L=4.0$) is metastable, the capsule seeks for a more favourable position by means of the random process of bond formation and dissociation. The equilibrium position in all the simulations of this study was associated with maximal average number of active bonds between the capsule and the wall, Fig.\ref{Fig-crawl-bonds}. The process is caused by the  stochastic binding to the distant pillars (over the distances $d > R$ from the center of mass of the capsule). This is allowed by the nature of the adhesive receptors, which can stretch due to thermal motion \cite{Mani2012}. Also, the adhesive microvilli of some living cells may extend due to dynamical growth of actin micro-filaments of their cytoskeleton \cite{Dembo1994, McConnell, Majstoravich}. These over-stretched bonds ($l>l_0$) may be established between a segment of the membrane and a distant pillar with a certain probability $P_{\rm on}$, according to the model Eq.(\ref{Pon}). Although their break-up probability $P_{\rm off}$ is close to $1$ and their lifetime is short, they nevertheless can cause a deformation of the vesicle making the formation of new bonds between this segment of the membrane and the distant pillar more favourable. The crawling stops if the distant pillars remain beyond the effective radius of bond formation. The latter decays exponentially with the distance. Hence the system demonstrates a destabilisation of the metastable state by random fluctuations caused by the dynamical attachment and breakage of adhesive bonds.

\section{Discussion and Conclusions}

In this study, a three-dimensional computational model for
adhesion of deformable capsules (vesicles) to micro-rough surfaces has been developed.
As a result of the ligand-receptor binding, the capsule gets deformed and pulled to the surface. The balance between  the elasticity of the membrane, the tension of the adhesive receptors and the surface relief determines the equilibrium shape of the adherent vesicle. 

The results can be extended to the analysis of the adhesion of spherical living cells (e.g. bacteria or white blood cells). The micro-roughness of the substrate (wall) results in a one of three possible adhesion modes. The first mode (``fakir'' state) is similar to the Cassie-Baxter state for a liquid drop on a textured (superhydrophobic) surface: the cell forms bonds only with the tops of the microrelief and remains almost undeformed. The second regime is the ``nested'' (contacting) state - when the cell membrane is deformed so that bonds can be formed with all the points of the surface relief while the membrane does not break. This adhesive state has been observed for relatively short micro-pillars (i.e. shallow relief). The third mode corresponds to the situation when the cell is pierced by a relief. This type of adhesion can lead to cell's death, it is realized by the pillars that are very thin compared to the spectrin mesh of the cytoskeleton of the cell. These results give new insights into  mechanisms of cell adhesion to surfaces with microscopic roughness. It has  also been found that the periodic array of pillars may cause a preferential localisation of the cells on the adhesive surface, depending on the $R/L$ ratio. This effect could be used for cell manipulation and positioning in vitro and also for cell sorting in microfluidics. 

The recommendations for the design of artificial antibacterial surfaces could be formulated on the basis of the adhesion maps obtained in the presented work: the impalement of the cell occurs if $r_p/L < 0.1$, thus narrow and dense pillars could in principle inhibit the formation of bacterial films. These findings could help in the design of a new generation of biomimetic materials for controllable and selective cell adhesion. 

The presented study has its limitations. Firstly, in the model the  ligand-receptor bonds are assumed to play the primary role in adhesion, while other non-specific forces are treated via a generalized repulsive force between the capsule and the wall. In reality electrostatic forces, excluded volume forces, hydrophobic/hydrophilic interactions may become significant in several cases and affect the adhesion. Secondly, it would be instructive to investigate the role of externally imposed fluid flows on the adhesion of cells or vesicles to surfaces with microscopic roughness. Hopefully, the present work will become a motivation for the experimental verification and further research.

\section*{Acknowledgements}

This work was partly supported by the Russian Foundation for Basic Research, project number 16-31-60061-mol-a-dk. The author also acknowledges the support from RUDN project 5-100. The simulations were carried out using computational resources of the Supercomputing Center of M. V. Lomonosov Moscow State University (``Lomonosov'' and ``Lomonosov-2'') \cite{Lomonosov}.

\section*{ Appendix: Mesh resolution test }

\begin{figure}
 \includegraphics[width=\columnwidth]{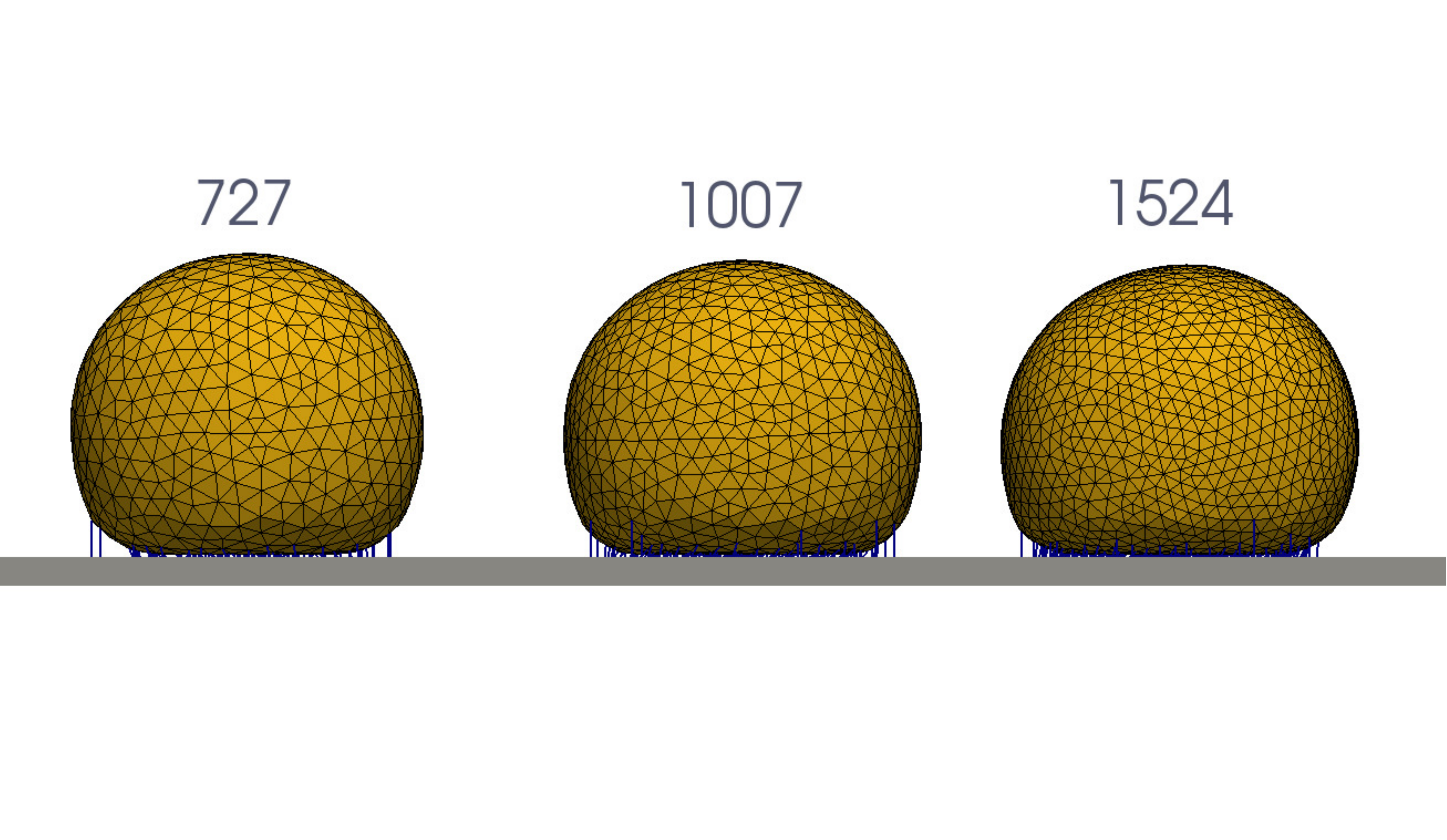}
  \caption{  The equilibrium shape of the adherent capsule for different mesh resolution: 727, 1007 and 1524 Lagrangian points }
\label{Fig-mesh-1}
\end{figure}

\begin{figure}
 \includegraphics[width=\columnwidth]{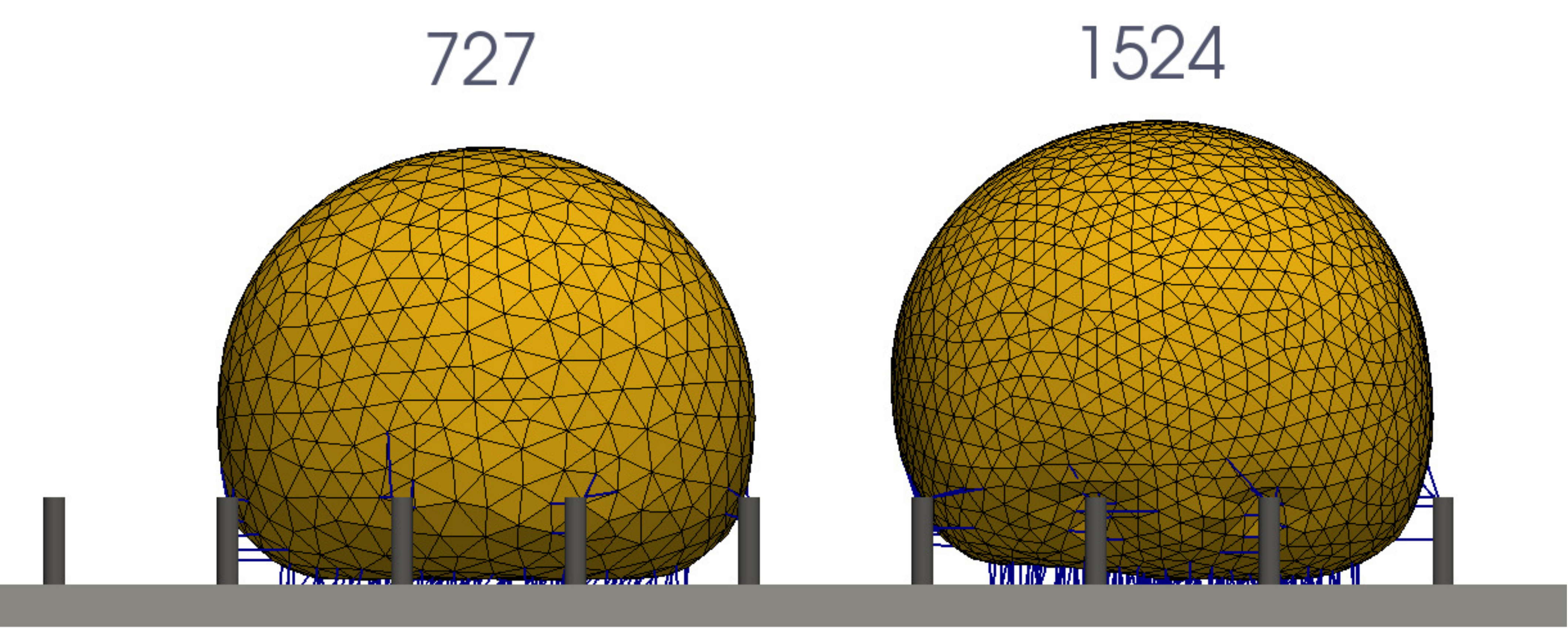}
  \caption{  The impaled state has been observed for the same parameters of the roughness at different mesh resolution: 727 (right) and 1524 (left) Lagrangian points. }
\label{Fig-mesh-2}
\end{figure}

To ensure that the simulation results are independent from the resolution of the mesh of Lagrangian points, from which the capsule has been constructed, a simulations for 1007 and 1524 LSPs have been conducted. Since the parameter $k_{\rm on}$ depends on the mesh density via the number of receptors per LSP ($n_r$), it had been scaled appropriately before each simulation. In general, there was no significant difference observed. The results are presented in Fig.\ref{Fig-mesh-1}. The impaled state has also been observed after the increase of the mesh resolution from 727 to 1524 LSPs, Fig.\ref{Fig-mesh-2}.

\bibliographystyle{plain}
\bibliography{adhesion}

\end{document}